\documentclass[aps,preprint, prb]{revtex4}
\usepackage{graphicx}
\usepackage{epsfig}
\usepackage{amsmath}
\usepackage{graphics}
\begin{document}

\title{Orbital interaction mechanisms of conductance enhancement and rectification by dithiocarboxylate
anchoring group}

\author{Zhenyu Li}
\author{D. S. Kosov }

\affiliation{Department of Chemistry and Biochemistry,
 University of Maryland, College Park, MD  20742}


\begin{abstract}
We study computationally the electron transport properties of dithiocarboxylate terminated 
molecular junctions. Transport properties are computed self-consistently 
within  density functional theory and non-equilibrium Green's functions formalism.
A microscopic origin of the experimentally observed current amplification by 
dithiocarboxylate anchoring groups is established.
We find that for the 4,4$^{\prime}$-biphenyl bis(dithiocarboxylate) junction the
 interaction of LUMO of the dithiocarboxylate anchoring  group with  LUMO and HOMO of the biphenyl part results
into bonding and antibonding resonances in the transmission spectrum in the  vicinity of the 
electrode Fermi energy.  A new microscopic mechanism of rectification is predicted based on the electronic structure of 
asymmetrical anchoring groups. We show that peaks in the transmission spectra of 
4$^{\prime}$-thiolato-biphenyl-4-dithiocarboxylate junction respond differently to the applied voltage.
Depending upon the origin of a transmission resonance 
in the orbital interaction picture, its energy
 can be shifted along with the chemical potential of the
electrode to which the molecule is stronger or weaker coupled. 
\end{abstract}


\maketitle

\section{Introduction}

One of the main goals in nanotechnology is the construction,
measurement and modeling of electronic circuits in which molecular
systems act as conducting elements.\cite{nitzan0384,joachim0501} When a
molecule is attached to two macroscopic metal electrodes with
different chemical potentials, the electric current flows through
it. 
The coupling to electrodes mixes the discrete molecular levels with the continuum of the metal electronic states such that 
 the molecular orbitals protrude deep inside the electrode. The coupling also renormalizes the energies of the molecular orbitals. Therefore, it is no longer correct to talk about the transport properties of the molecule,
 but rather, only of the  electrode-molecule-electrode heterojunction. If the strength of the
 molecule-electrode coupling is large, substantial perturbation of the molecular electronic structure can occur. In fact,
 it is expected that upon initial chemisorption, substantial charge transfer takes place between the 
 metal electrode and the molecule even in the absence of an applied voltage bias.   
These effects are pivotal for molecular wire transport properties and they can be controlled by changing the interface geometry or by altering the anchoring groups, which  provide the linkage between the molecule and the 
electrode.\cite{venkataraman0600, li0635, kim0670, xue0306,ke0497, basch0575, ke0501, li06xx} 
Therefore the study of the anchoring group chemistry in molecular electronics may help us not only to pin down the
origin of  significant discrepancy between experimental and theoretical molecular conductivities but may also reveal  new fascinating fundamental aspects of molecule-surface interactions. 
With only a few exceptions,\cite{siaj0588, tulevski0591, guo0656}  the most widely used molecular wire
junctions have been formed so far by organic molecules assembled between  gold electrodes via thiol anchoring groups. However,  thiol linkage is considered to be only structural and lacks any subsequent useful "chemistry",\cite{tulevski0591} since the energy and the electron occupation of sulfur 3$p$ orbital is difficult to modify. Tulevski \textit{et al.} \cite{tulevski0591} thus suggested to use ruthenium electrode instead of gold. But being excellent conductor and metal with mature manipulation techniques, gold remains the most attractive electrode material.

The quest for reliable molecular electronic devices has become the search for better molecule-gold 
linkers,\cite{akkerman0669} which
provide  the opportunity  not only to grow  the single molecular contacts  but also to control molecular transport
properties. Recently, Tivanski \textit{et al.}\cite{tivanski0598} suggested and realized experimentally the
remarkable molecular wire, which is attached to the gold electrodes via dithiocarboxylate conjugated linker (-CS$_2$). The
conductance of 4,4$^{\prime}$-biphenyl dithiolate (BDT) and 4,4$^{\prime}$-biphenyl
bis(dithiocarboxylate) (BDCT) molecular wires were measured by
conducting-probe atomic force microscopy.\cite{tivanski0598}  The central conducting
parts of both BDT and BDCT are exactly the same (biphenyl) whereas
the anchoring groups are different. BDT has the standard
thiol groups and BDCT is terminated by dithiocarboxylate groups.
  It was experimentally observed that the
conductance of BDCT is 1.4 times as large as that of BDT.\cite{tivanski0598} 
But the physical origin of this conductance enhancement was not
clear. The most intuitive picture suggests  that  the conjugated dithiocarboxylate anchoring 
group provides  the stronger coupling between the electrode and the molecule.\cite{tivanski0598} We have recently demonstrated that this  simple mechanism
plays a central role in conductance enhancement induced by dithiocarbomate linker (N-CS$_2$),\cite{li06xx} where 
the stronger molecule-electrode coupling  leads to the larger mixing between the discrete molecular levels and the continuum of the
metal electronic states, and thus to the 
larger broadening of resonances in the electron transmission spectrum. As it turned out  
 the mechanism of the conductance enhancement is entirely 
different for dithiocarbamate and dithiocarboxylate linkers, although they are structurally very similar.
One of the aims of this paper is to elucidate the  microscopic origin of the conductance enhancement via dithiocarboxylate linkers. 
We show that the reason for the conductance enhancement is not simply the difference in the
molecule-electrode coupling strengths but the disparity in the electronic structure of  thiol and  dithiocarboxylate anchoring groups. 

One of the interesting possibilities, which we also would like to explore  in our paper, is the use of  dithiocarboxylate 
linkers to create a molecular rectifier.  A molecular rectifier is  a junction where electrons flow along one preferential direction.\cite{aviram7477,metzger0303} As a prototype molecular junction we consider 
biphenyl with thiol linker on one side and with dithiocarboxylate anchoring group on the 
other -- 4$^{\prime}$-thiolato-biphenyl-4-dithiocarboxylate -- (TBCT).
Rectification for molecules with asymmetric anchoring groups has been studied 
theoretically,\cite{taylor0201, li06xx} however, in all junctions considered so far the role of anchoring groups was 
limited to providing left-right asymmetry in the coupling strength between the molecule and the electrodes.
It leads to the standard coupling-strength picture of molecular rectification, which predicts that as the voltage bias increases
 the peaks in the transmission are shifted along with the chemical potential of the electrode to which the molecule
 is stronger coupled.\cite{taylor0201, li06xx} Suppose that, for example, there are two resonances 
 in the transmission spectrum in the vicinity of the electrode Fermi energy  with the energies above  and below it, so that  both resonances contribute  to the electron transport at moderate voltages. Within the standard rectification picture the two peaks are shifted in the same direction.\cite{taylor0201, li06xx}
 Therefore, when the energy of one resonance enters the integration range between  the chemical potentials 
 of the left and the right electrodes
 (i.e. it starts to contribute to the electron current)
 the second resonance could be already shifted away from the integration range. Thus, 
 as soon as  the contribution from  one peak to the current increases, the role of the second pick decreases. This counterbalancing reduces  the rectification effects within the standard coupling-strength mechanism.
  In this paper we show that the TBCT molecular junction
  exhibits an entirely new mechanism of rectification, which overcomes the limitations of the standard 
  coupling strength picture described above.
%

The remainder of the paper is organized as follows. In section II,
we describe the computational details. The main results are
discussed in section III. We first illustrate how the electronic
structure of  thiol and  dithiocarboxylate anchoring groups
control and determine transport properties of molecular wire
junctions. A new rectification mechanism is  suggested for 
molecule attached to  electrodes by thiol and dithiocarboxylate groups. Section IV concludes the paper

\section{Computational methods}
Two computer programs were used in our calculations. First,  optimized geometries of the molecular junctions were obtained
by SIESTA computer program.\cite{soler0245}  Then
electron transport properties were computed by using TranSIESTA-C 
package. \cite{brandbyge0201}  TranSIESTA-C uses the combination of 
non-equilibrium Green's function (NEGF) formalism and  density
functional theory (DFT). In NEGF theory, the molecular wire
junction is divided into three regions: left electrode (L),
contact region (C), and right electrode (R). The semi-infinite electrodes are calculated separately to obtain
the bulk self-energy.
The contact region
contains parts of the electrodes to include the screening  effects in the calculations.
 The electrodes are modeled by semi-infinite Au surfaces.
The main loop for TranSIESTA-C self-consistent NEGF/DFT calculations is described below (for technical details
we refer to paper\cite{brandbyge0201}).
The matrix product of the Green's function and the imaginary part of the left/right electrode self-energy yields the spectral densities. The spectral densities of the left and right electrodes are combined together to compute the nonequilibrium, 
voltage-dependent density matrix and then the density matrix is converted into nonequilibrium electron density.
The nonequilibrium electron density enables us to compute matrix elements of Green's function. The Hartree potential is determined through the solution of the Poisson equation with appropriate voltage-dependent boundary conditions. 
This loop of calculations is repeated until self-consistency is achieved.
After the self-consistent convergence is
achieved, the voltage transmission spectrum is calculated by the standard equation
\begin{equation}
\label{eq:trans} T(E, V)=Tr[\mathbf{\Gamma}_L(E, V)\mathbf{G}(E,V)\mathbf{\Gamma}_R(E, V)\mathbf{G}^\dagger(E, V)],
\end{equation}
where $\mathbf{G}$ is the Green's function of the contact region,
$\mathbf{\Gamma}_{L/R}$ is the coupling matrix, and  $V$ is the applied voltage bias.
The electric current as a function of the applied voltage is obtained  by the
integration of the transmission spectrum.
\begin{equation}
\label{eq:iv} I(V)=(-e)\int T(E,V) (f(E-\mu_L) -f(E-\mu_R))dE,
\end{equation}
where $f$ is the Fermi-Dirac occupation number,  $\mu_L =-eV/2$ ($\mu_R=eV/2$) is the chemical potential
for the left (right) electrode and $e$ is the elementary charge.

 We use double-$\zeta$ with polarization (DZP) basis  for all atoms except Au, for which single-$\zeta$ with polarization (SZP) is used. We use Troullier-Martins nonlocal pseudopotentials in all our calculations.\cite{troullier9193} The exchange-correlation potential is described by Perdew-Zunger 
local density approximation (LDA).\cite{perdew8148}
Single $k$-point sampling on the plane perpendicular to the direction of the current is used in our calculations. Our tests show that the generalized gradient approximation\cite{perdew9665}  yields negligible corrections to the LDA transmission spectra and 
the use of 4 $\times$ 4 $k$-point sampling does not affect main features of the transmissions.

\section{Results and Discussion}
\subsection{Structural model of molecular junctions}
\begin{figure}
\includegraphics[keepaspectratio,totalheight=10cm]{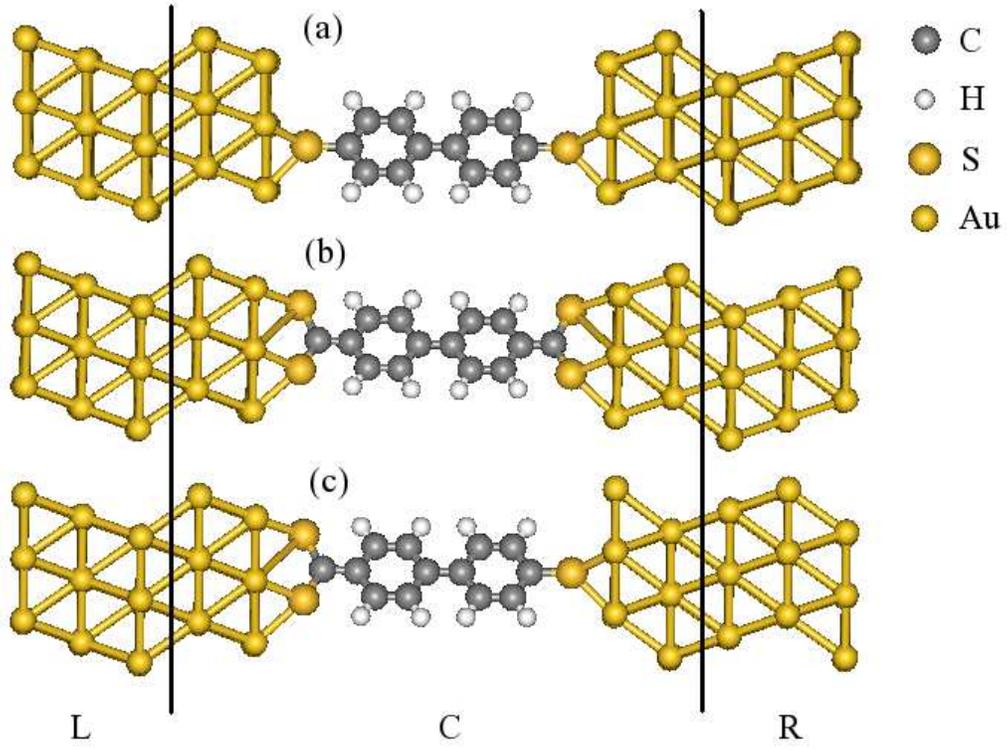}
\caption{ 
Relaxed geometry of molecular wire 
junctions with different anchoring groups. (a) BDT, (b) BDCT, (c)
TBCT. Only one unit cell for the semi-infinite electrode is
plotted.
} \label{fig:geo}
\end{figure}

We consider three representative molecular wire junction systems:
BDT, BDCT, and TBCT. BDT and  BDCT are terminated by  thiol and dithiocarboxylate 
respectively on both ends. In TBCT, dithiocarboxylate group is used on the left side 
while thiol group is used on the right side. Figure \ref{fig:geo} shows the
optimized junction geometries. The semi-infinite left and right electrodes are modeled by two Au(111)-(3$\times$3) surfaces,
for which only one unit cell (contains three Au layers) is plotted. In the contact region, two Au layers at both left and
right side are included. The outmost left and right layers in the contact regions are constrained to their theoretical bulk 
geometry to match the structure of  Au(111)-(3$\times$3) surfaces which are used to model the electrodes. We assume the  biphenyl interior part of all three junctions have coplanar geometry.  
This narrows down the problem of difference between BDT, BDCT and TBCT transport properties
to the role of the dithiocarboxylate linkers. The rest of the contact region is fully optimized. We also optimized the length of the junctions
by computing  the total energies of the systems as functions of the distance between the left and the
right electrodes. Every single energy point  is calculated by performing  geometry optimization with constrained
electrode-electrode separation. The optimal separation between the electrodes is obtained as the distance at which
the total energy is minimal. We also perform test calculations for non-coplanar BDT and BDCT.  
The twisting of phenyl rings by $37^\circ$, which corresponds to the equilibrium geometry for 
biphenyl junction,\cite{xue0306}  has negligible  influence on the transmission spectra. The twisting merely reduces the overall 
transmission probability without any further alternation of the transmission spectra.

\subsection{Orbital interaction mechanism of conductance enhancement}
\begin{figure}
\includegraphics[keepaspectratio,totalheight=10cm]{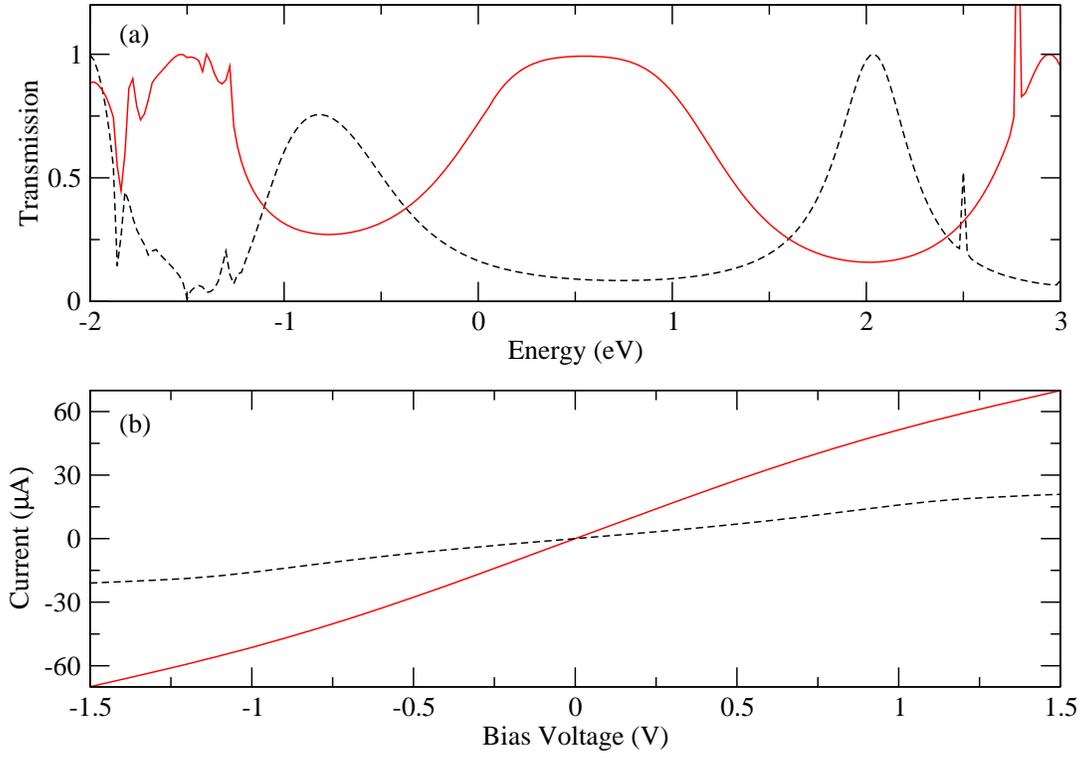}
\caption{
(a) Zero voltage bias transmission spectra and
(b) current-voltage characters of BDT (dashed) and BDCT
(solid) molecular  junctions. Fermi energy of the
electrode is set to zero in the transmission spectra.
}
\label{fig:trans}
\end{figure}

With the optimized geometries, we calculate  transmission spectra and current-voltage (I-V) characteristics for 
BDT and BDCT, as shown in Figure \ref{fig:trans}. The computed current is four orders of magnitude 
larger than the experimental one.\cite{tivanski0598}
 Our I-V curves qualitatively reproduce 
the experimental conductance enhancement although theoretically predicted  
increase in conductivity ($\sim$ 2.1) is larger than the experimentally observed amplification factor ($\sim$ 1.4).\cite{tivanski0598}
The conductance enhancement is  clearly
shown in the transmission spectra. For BDT, there are two broad peaks: one is below the electrode 
Fermi energy ($\sim$ -0.8 eV) and the other is above it ($\sim$ 2.1 eV). These two peaks
can still be found in the transmission spectrum of BDCT, but the
first peak is shifted toward the lower energy ($\sim$ -1.5 eV) and the second peak
is moved to the higher energy ($\sim$ 2.9 eV). Besides these two peaks, the additional
broad peak appears for BDCT just above the Fermi energy. It gives almost perfect transmission probability 
in the broad energy range from 0.1 to 1.1 eV. This broad peak results in large
transmission probability at the Fermi energy and  is responsible for the conductance enhancement for  BDCT junction
observed on experiment. \cite{tivanski0598}

\begin{figure}
\includegraphics[keepaspectratio,totalheight=14cm]{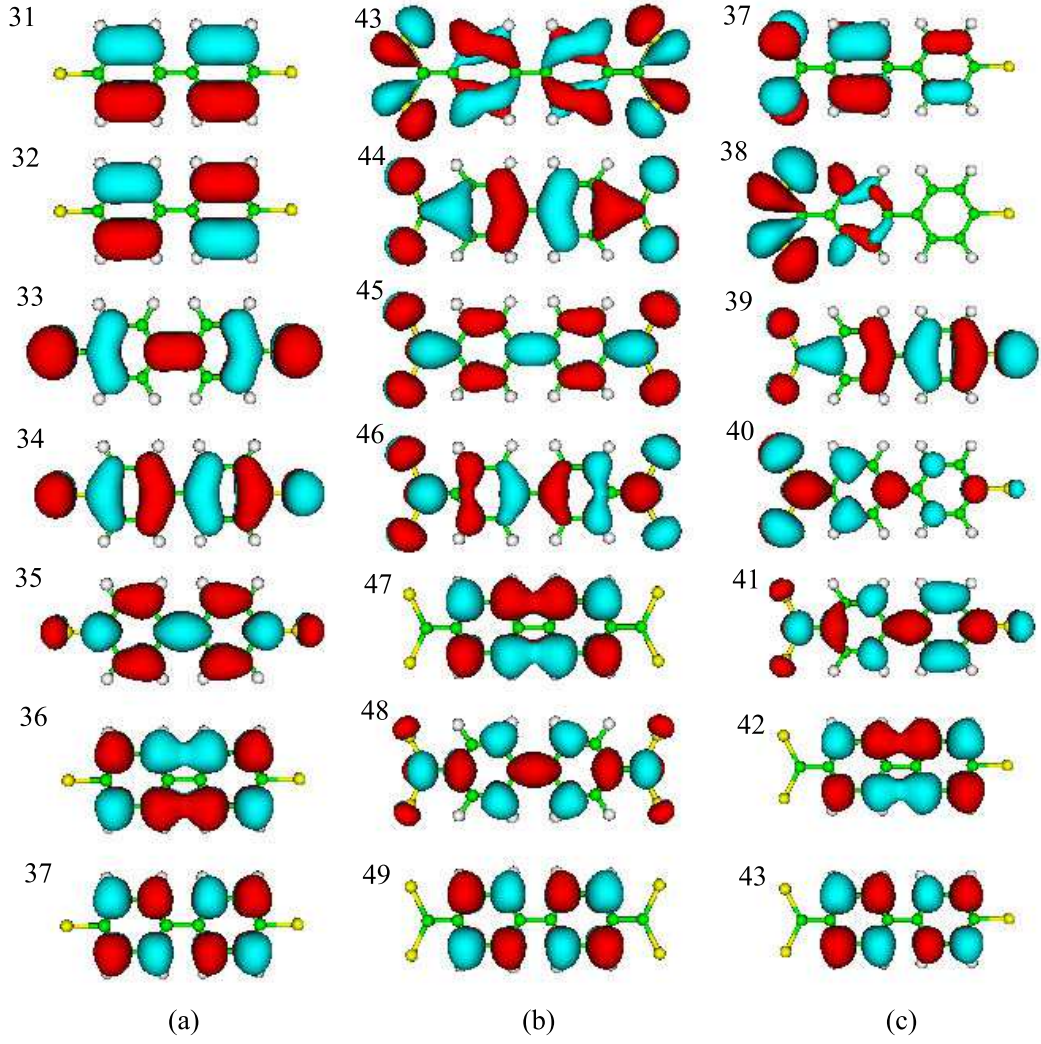}
\caption{ 
MPSH  orbitals near gold electrode Fermi
level for (a) BDT, (b) BDCT, and (c) TBCT. 
} \label{fig:MPSH}
\end{figure}

The microscopic origin of this broad peak and thus the cause of the conductance
enhancement can be understood from the partitioning of the transport channels into contribution from the interior part
(biphenyl) and the linkers (thiol or dithiocarboxylate). If we project the self-consistent hamiltonian onto the Hilbert space spanned by
the basis functions of the molecule (includes biphenyl and anchoring groups), we obtain the molecular projected
self-consistent hamiltonian (MPSH). The eigenstates of MPSH can be
considered as molecular orbitals renormalized by the
molecule-electrode interaction. Figure \ref{fig:MPSH} shows the MPSH
orbitals near the Fermi level of the electrodes and Table
\ref{tbl:energy} shows the corresponding eigenvalues. Comparing
the energies of the resonances in the transmission spectra with the
eigenvalues of MPSH, we find that the two main transmission peaks of
BDT are mainly contributed by MPSH orbitals 34 and 35, while the
peaks of BDCT take their origin from MPSH orbitals 44, 45, 46, and
48. The common feature of these orbitals is the significant
de-localization and spread of the electron density along the
interior region of the wire as well as the anchoring groups. The
additional peak in the transmission spectrum of BDCT is not a single
broad resonance. As is elucidated by the MPSH analysis, the merging of
two resonances from MPSH orbitals 45 and 46 produces this broad
peak. Peaks in BDTC transmission do not show significant additional 
broadening with respect to  BDT resonances.

\begin{figure}
\includegraphics[keepaspectratio,totalheight=10cm]{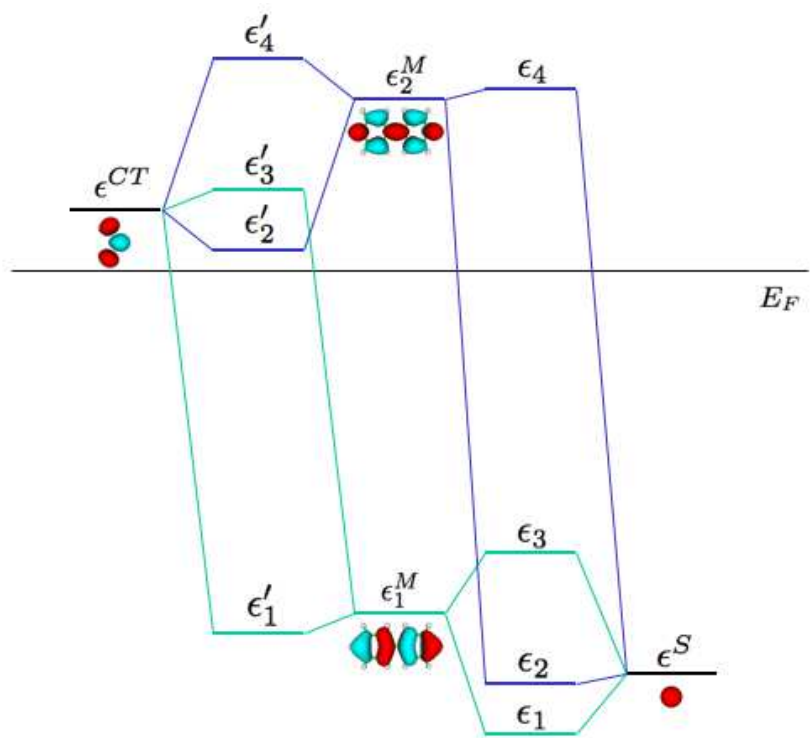}
\caption{
Schematic orbital interaction picture. 
Interaction between biphenyl HOMO/LUMO and anchoring group orbitals (sulphur $p$ and dithiocarboxylate LUMO)
leads to MPSH orbitals.
}
\label{fig:levels}
\end{figure}

The MPSH orbitals can be further disentangled if we consider them
as generated by the interaction between orbitals localized on the
anchoring groups and orbitals of the interior part (biphenyl
molecule). Dissecting all relevant MPSH orbitals for the three
junctions shows that all of them can be obtained from  linear
combinations of the two anchoring group orbitals (sulphur  3$p$ orbital
of thiol linkage and LUMO of dithiocarboxylate) and the HOMO/LUMO
of biphenyl molecule. This orbital interaction picture is
presented in Figure \ref{fig:levels}. The biphenyl HOMO and LUMO
(orbital 28 and 29 in Table  \ref{tbl:energy}),  are labeled as
$\epsilon^M_1$ and $\epsilon^M_2$ in Figure \ref{fig:levels}.
 Sulfur 3$p$ atomic orbital of thiol linkage
($\epsilon^S$) lays very deep (-2.88 eV from calculation of atomic energy levels of sulphur) below the Fermi energy of
the gold surface.\cite{tulevski0591} When thiol group is attached
to biphenyl, the resulted MPSH orbitals are all far from the Fermi
level, and only the anti-bonding orbitals ($\epsilon_3$ and
$\epsilon_4$) contribute to low voltage electron transport. From
Fig. \ref{fig:MPSH}a, we can easily identify  $\epsilon_3$ and
$\epsilon_4$ as BDT MPSH orbitals 34 and 35 respectively.

The orbital interaction picture leads to  very different mechanism of  electron transport 
through  BDCT molecular wire. Here,  orbital interaction between the
LUMO of the dithiocarboxylate group ($\epsilon^{CT}$) and the
biphenyl HOMO/LUMO contributes to the four resonance structures in
the transmission spectrum. The value of $\epsilon^{CT}$ (0.35 eV  from calculation of H$_2$S$_2$--CH molecular orbitals)
is much higher than $\epsilon^{S}$ in the thiol, and, therefore,
both bonding and anti-bonding MPSH orbitals of BDCT are positioned
in the vicinity of the Fermi energy and contribute to low bias
electron transport. The right side of Figure \ref{fig:levels} shows the
orbital interaction picture for BDCT, where levels
$\epsilon_1^{\prime}$ to $\epsilon_4^\prime$ correspond to MPSH
orbital 44, 45, 46, and 48 respectively. Level
$\epsilon_2^{\prime}$, which is almost in resonance with the
electrode Fermi energy, gives the main contribution to  low
voltage electron transport. Therefore, it is the intrinsic
electronic structure difference between the anchoring groups that
is the real origin for the observed conductance enhancement by the
dithiocarboxylate anchoring group.

\subsection{Rectification}
In this section, we demonstrate that the biphenyl molecular wire with thiol linker on one side and dithiocarboxylate anchoring
group on the other (TBCT) exhibits a new mechanism of rectification. The mechanism  enables us to overcome 
the limitation of the  standard coupling-strength picture,\cite{taylor0201, li06xx}
in which rectification from one resonance could be reduced by the opposite contribution from another resonance.

\begin{figure}
\includegraphics[keepaspectratio,totalheight=14cm]{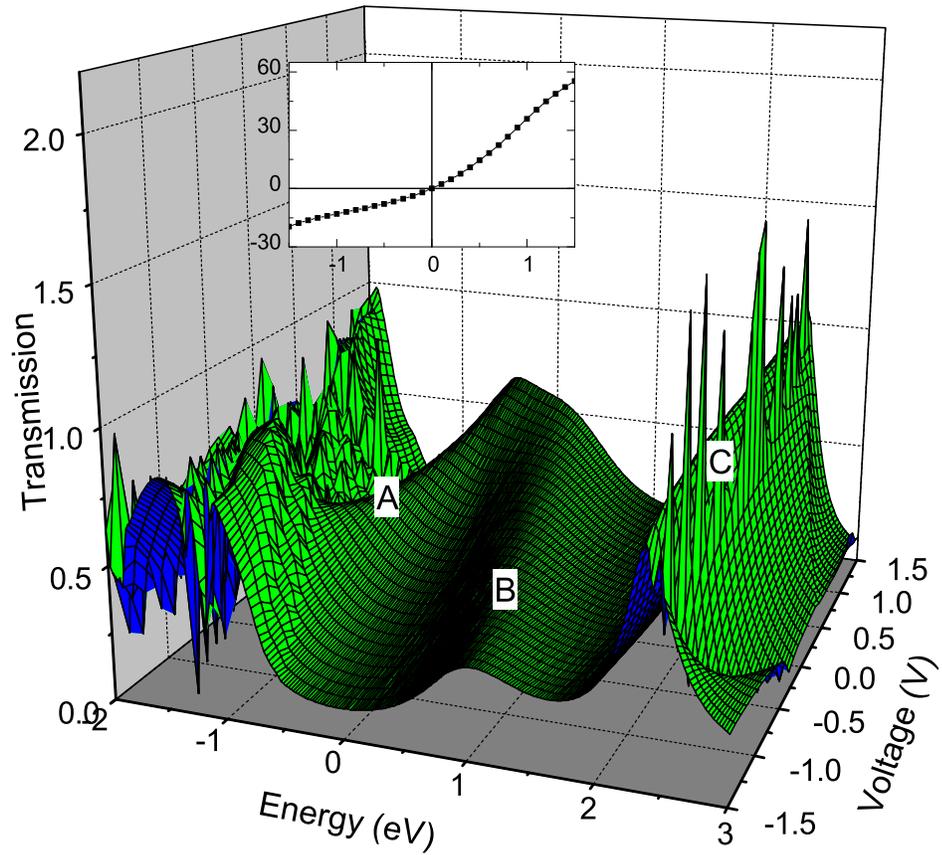}
\caption{
Voltage-dependant  transmission spectrum of the TBCT junction.
Inset: current-voltage curve with voltage (in V) as the abscissa and
current (in $\mu$A) as the ordinate.
}
\label{fig:asymm}
\end{figure}

The voltage dependence of the transmission spectrum of  TBCT junction  shows very complicated patterns (Figure \ref{fig:asymm} and
Table \ref{pk}).
There are three main transmission peaks. The left peak (A) shifts toward the higher energy as the voltage changes from
negative to positive bias, the middle peak (B) shows the opposite behavior, and the right peak (C) does not shift at all. It means
that peak A follows the changes in the chemical potential of the right electrode ($\mu_R=eV/2$), while peak B follows the chemical
potential of the left electrode ($\mu_L=-eV/2$). The current-voltage characteristics is shown in the inset of Figure \ref{fig:asymm}. It is obtained by integrating the transmission from $\mu_L$ to $\mu_R$. For negative bias voltage, both peaks A and B are
not within the integration range, therefore the current is small. Under positive voltage, when the value of the voltage increases, both A
and B enter the integration region and produce large current increase. We thus obtain rectification coefficient 
$R=I(V)/I(-V)\sim 2.8$ at 1.0 eV bias voltage. 

The response of the transmission spectrum on the applied voltage bias can be readily understood if we look at the MPSH orbitals as plotted in Figure \ref{fig:MPSH}c. Peaks A, B, and C correspond to MPSH orbitals 39, 40, and 41 respectively. Orbital 39 has stronger
molecule-electrode coupling via the thiol group so that the energy of this orbital follows the chemical potential of the right
electrode. This orbital is mainly antibonding mixture of the HOMO of biphenyl ($\epsilon^M_1$ in Fig. \ref{fig:levels}) 
and $3p$ orbital of the sulphur ($\epsilon^S$ ).
Therefore  orbital 39  
($\epsilon_3$ plus minor contribution from $\epsilon^{\prime}_1$) 
has stronger coupling at the thiol side. Similarly, orbital 40 is mainly related to
$\epsilon_2^\prime$, which has stronger coupling via the
dithiocarboxylate group and follows the chemical
potential of the left electrode. Orbital 41 is anti-bonding at
both sides, and can be considered as combination of $\epsilon_4$
and $\epsilon_4^\prime$. It interacts equally well with both
electrodes. Therefore the influence of the left electrode is
balanced by the influence of the right electrode and this orbital
is not significantly affected by the applied voltage. 
To summarize, likewise in the case of the conductance enhancement, the electronic structure of the anchoring groups 
and orbital interaction picture provide the explanation why these three transmission peaks behave differently 
under the voltage  bias.

\begin{figure}
\includegraphics[keepaspectratio,totalheight=10cm]{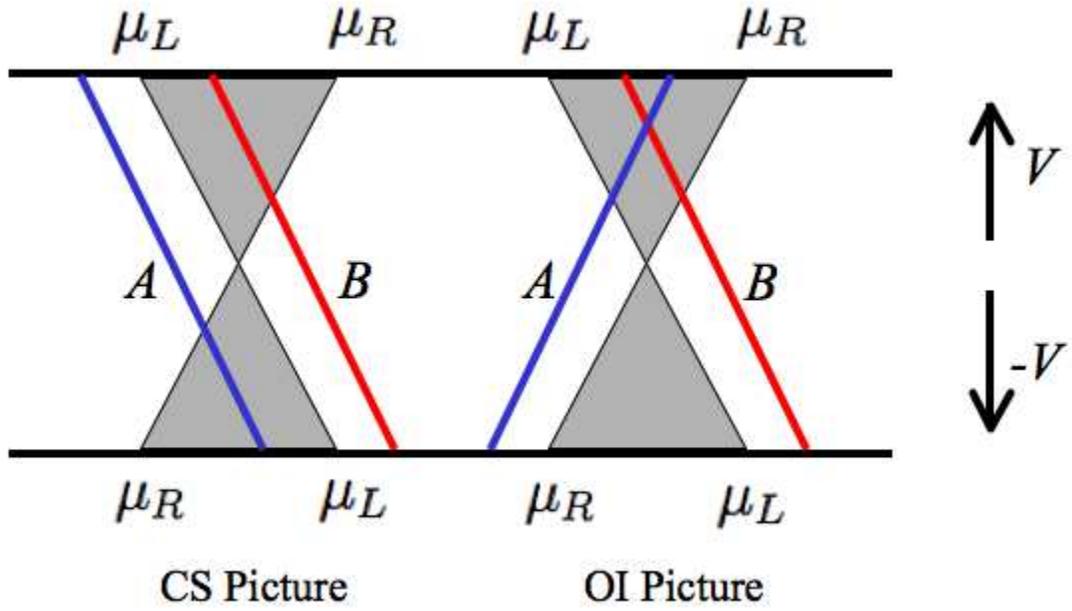}
\caption{ Coupling
strength (CS)  and orbital interaction (OI) mechanisms for current rectification.  
The transmission from the shadow region contributes  to current (see Equation (\ref{eq:iv})). 
Blue and red lines represent two transmission
resonances  $A$ and $B$.} \label{fig:rect}
\end{figure}

 Schematic diagram which shows the qualitative behavior of the two peaks is presented 
on Figure \ref{fig:rect}. In the standard coupling-strength picture, all transmission peaks follow the chemical potential of the stronger coupled electrode (left electrode in the figure).\cite{taylor0201, li06xx} If there are two transmission peaks equally spaced below and above the mean Fermi level of the electrode (A and B on Figure \ref{fig:rect})  their contributions to the rectification counteract each other. 
At negative bias, peak A enters the integration range, which  increases the current comparing to that at the positive bias. 
The current  prefers to flow from the  right to  the left due to peak A.  On the other hand, peak B increases the current at positive bias voltage, and, therefore the current prefers to flow the left to right because of peak B. This situation can be avoided in
the  orbital interaction picture, where peaks A and  B respond oppositely on the applied voltage.
As shown in the right part of Figure \ref{fig:rect}, which is exactly the situation we 
 find in the TBCT junction,  where both peaks A and B rectify the current in the same direction.

\section{Conclusions}

We have  performed DFT-NEGF calculations to elucidate the  microscopic origin of the conductance enhancement via dithiocarboxylate linkers. 
We showed that the reason for the conductance enhancement is not simply the difference in the
molecule-electrode coupling strengths but the disparity in the electronic structure of  thiol and  dithiocarboxylate anchoring groups. 
We suggested the use of dithiocarboxylate linker to create a molecular rectifier.  As a prototype molecular rectifier we considered
 a biphenyl with a  thiol linker on one side and with  a dithiocarboxylate anchoring group on the 
other (TBCT molecule). We predicted that the TBCT molecular junction exhibits an   entirely new mechanism of rectification which had never been predicted theoretically or observed experimentally before. 
Our calculations demonstrate how electronic structure of anchoring groups accompanied 
by molecular orbital interaction picture can be used as a guiding principle to predict transport 
properties of molecular junctions.

\begin{acknowledgments}

The authors are grateful to M. Gelin, L. Sita, A. Vedernikov, and
J. Yang for helpful discussion. This work was partially supported by the American Chemical Society Petroleum Research Fund (44481-G6) and by  summer award of General Research Board of the University of Maryland. 
\end{acknowledgments}

\vspace{3cm}
{\bf Supporting Information Available }

Total  energies of the molecular junctions as functions of electrode-electrode separation, transmission spectra for different molecular geometries, transmission spectra computed with GGA exchange-correlation functional, and  transmission spectra computed with 4$\times$4 $k$ sampling.  This material is available free of charge via the Internet at http://pubs.acs.org.

\clearpage

\begin{table}[b]
\caption{ Energies of molecular orbitals of biphenyl and MPSH
orbitals of BDT, BDCT, and TBDT. All energies are relative to the
Fermi energy of the gold electrode.} \label{tbl:energy}
\begin{tabular}{ccccccccccc}
\hline\hline
\multicolumn{2}{c}{Biphenyl} &&  \multicolumn{2}{c}{BDT} && \multicolumn{2}{c}{BDCT}  && \multicolumn{2}{c}{TBDT} \\
\cline{1-2}\cline{4-5} \cline{7-8}\cline{10-11}
$n$ & $\epsilon_n$(eV) && $n$ & $\epsilon_n$(eV) &&  $n$ & $\epsilon_n$(eV) &&  $n$ & $\epsilon_n$(eV)   \\
\hline
25 & -3.926 && 31 & -2.580 && 43 & -2.222 && 37 & -2.343 \\
26 & -3.201 && 32 & -2.433 && 44 & -1.821 && 38 & -2.230 \\
27 & -3.052 && 33 & -2.341 && 45 &  0.003 && 39 & -1.489\\
28 & -2.358 && 34 & -1.271 && 46 &  0.849 && 40 &  0.433 \\
29 &  1.229 && 35 &  1.838 && 47 &  2.772 && 41 &  2.305 \\
30 &  1.840 && 36 &  2.506 && 48 &  2.808 && 42 &  2.559 \\
31 &  2.350 && 37 &  3.020 && 49 &  3.257 && 43 &  3.087  \\
\hline\hline
\end{tabular}
\end{table}

\clearpage

\clearpage

\clearpage

\clearpage

\clearpage

\begin{table}
\caption{Energies  (in eV) of  the peaks A, B and C from Figure \ref{fig:asymm} for different values  of the applied voltage (in V). } 
\label{tbl:table1}
\begin{tabular}{ccccccc}
\hline\hline
voltage &&       A  &&  B  && C \\
\hline
1.5  &&   - 0.4  && 0.0  &&  2.4\\
0.5  &&  -0.7   &&  0.3 &&  2.4 \\
0.0  && -0.9  && 0.5  && 2.4 \\
-0.5 &&   -1.3  &&  0.7 && 2.4 \\
-1.5 &&  -1.5   &&  1.0 &&  2.4
\\
\hline
\hline
\end{tabular}
\label{pk}
\end{table}

\clearpage

\end{document}